\documentclass{emulateapj}

\shorttitle{NSMs as Origin of \lowercase{$r$}-Elements}
\shortauthors{Ishimaru et al.}

\begin{document}
%%======================================================================

%%=== Title ===
\title{Neutron Star Mergers as the Origin of \lowercase{$r$}-Process
Elements in the Galactic Halo 
Based on the Sub-halo Clustering Scenario}

%%=== Authors ===
\author{Yuhri Ishimaru\altaffilmark{1,3},
        Shinya Wanajo\altaffilmark{2},
        and 
        Nikos Prantzos\altaffilmark{3}
        }

\altaffiltext{1}{Department of Material Science,
        International Christian University,
        3-10-2 Osawa, Mitaka, Tokyo 181-8585, Japan;
        ishimaru@icu.ac.jp}

\altaffiltext{2}{iTHES Research Group, RIKEN, 
       2-1 Hirosawa, Wako, Saitama 351-0198, Japan;
        shinya.wanajo@riken.jp}

\altaffiltext{3}{Institut d'Astrophysique de Paris, UMR7095 CNRS,
        Univ. P. \& M. Curie, 98bis Bd. Arago, 75104 Paris, France;
        prantzos@iap.fr}

%%=== Abstract ===
%%
\begin{abstract}
Binary mergers (NSMs) of double neutron star (and black hole--neutron
star) systems are suggested to be major sites of $r$-process elements in
the Galaxy by recent hydrodynamical and nucleosynthesis studies. It has
been pointed out, however, that the estimated long lifetimes of neutron
star binaries are in conflict with the presence of $r$-process-enhanced
halo stars at metallicities as low as [Fe/H] $\sim -3$. To resolve this
problem, we examine the role of NSMs in the early Galactic chemical
evolution on the assumption that the Galactic halo was formed from
merging sub-halos. We present simple models for the chemical evolution
of sub-halos with total final stellar masses between $10^4\, M_\odot$
and $2 \times 10^8\, M_\odot$. Typical lifetimes of compact binaries are
assumed to be 100~Myr (for 95\% of their population) and 1~Myr (for
5\%), according to recent binary population synthesis studies.
The resulting metallcities of sub-halos and their ensemble are
consistent with the observed mass-metallicity relation of dwarf galaxies
in the Local Group, and the metallicity distribution of the Galactic
halo, respectively.
We find that the $r$-process abundance ratios [$r$/Fe] start increasing
at [Fe/H] $\le -3$ if the star formation efficiencies are smaller for
less massive sub-halos.
In addition, the sub-solar [$r$/Fe] values (observed as [Ba/Fe] $\sim
-1.5$ for [Fe/H] $< -3$) are explained by the contribution from the
short-lived ($\sim 1$~Myr) binaries. Our results indicate that NSMs may
have a substantial contribution to the $r$-process element abundances
throughout the Galactic history.
\end{abstract}

%%=== Keywords ===
%%
\keywords{
nuclear reactions, nucleosynthesis, abundances
--- Galaxy: evolution 
--- Galaxy: halo 
--- galaxies: dwarf 
--- stars: abundances
--- stars: neutron 
}

%%======================================================================
\section{Introduction}
%%======================================================================

The astrophysical origin of the rapid neutron-capture ($r$-process)
elements is still unknown. Recent spectroscopic studies of Galactic halo
stars have revealed the presence of Eu (almost pure $r$-process element
in the solar system) in the atmospheres of extremely metal-poor (EMP)
stars down to [Fe/H]
\footnote{$[A/B] \equiv \log (N_A/N_B) - \log (N_A/N_B)_\odot$, where
$N_A$ indicates the abundance of $A$.} $\sim -3$ with a large
star-to-star scatter in [Eu/Fe] \citep[more than 2 orders of magnitude,
e.g.,][]{Honda2004, Francois2007, Sneden2008}.
These observational features are reasonably reproduced by models of
\textit{inhomogeneous} Galactic chemical evolution (GCE), where a
limited mass range (e.g., $8-10 M_\odot$) of core-collapse supernovae
(CCSNe) is assumed to be the dominant source of the $r$-process elements
\citep{Ishimaru1999, Tsujimoto2000, Ishimaru2004, Argast2004}. Previous
\textit{homogeneous} GCE models that assume a well-mixed, one-zone
Galactic halo also suggest the low-mass end of CCSNe as the major
$r$-process site \citep{Mathews1992, Travaglio1999} but are not capable
of accounting for the star-to-star scatter in [Eu/Fe].

To date, however, no study of CCSN nucleosynthesis satisfactorily
accounts for the production of heavy $r$-process elements \cite[see,
e.g.,][]{Wanajo2011, Wanajo2013}. As an alternative scenario, binary
mergers of double neutron stars (NSMs) as well as of neutron star--black
hole pairs have long been suggested to be the $r$-process sites
\citep{Lattimer1974, Lattimer1976, Lattimer1977, Symbalisty1982,
Eichler1989, Meyer1989}. In fact, a number of recent nucleosynthesis
studies based on hydrodynamical simulations support NSMs to be the
promising sources of $r$-process elements in the Galaxy
\citep{Freiburghaus1999, Goriely2011, Korobkin2012, Bauswein2013,
Rosswog2014, Wanajo2014}.

Recent GCE models appear to disfavor the NSM scenario except for the
cases invoking very short lifetimes of neutron star binaries
\citep[1--10~Myr,][]{Argast2004, DeDonder2004, Matteucci2014,
Komiya2014, Tsujimoto2014, vandeVoort2015, Wehmeyer2015}
\footnote{ Note that the higher resolution model of
\citet{vandeVoort2015} also suggests short lifetimes of neutron star
binaries.  }.
Estimates of binary lifetimes, $t_\mathrm{NSM} \sim 0.1$--1~Gyr
\citep[e.g.,][]{Dominik2012}, are in conflict with the presence of
$r$-process-enhanced stars below [Fe/H] $\sim -2.5$ in those
models. Even if a short-lived channel with $t_\mathrm{NSM} \sim 1$~Myr
exists \citep{Belczynski2001, DeDonder2004}, the low Galactic rate of
such events \citep[0.4--77.4~Myr$^{-1}$,][]{Dominik2012} leads to a too
large star-to-star scatter in [Eu/Fe] at [Fe/H] $\gtrsim -2.5$
\citep{Qian2000, Argast2004}. It should be noted, however, that the
inhomogeneous GCE model of \citet{Argast2004} also appears to be in
conflict with the observed small scatter of $\alpha$ elements relative
to iron \citep{Argast2002}. This problem would be cured if the
inhomogeneity of the inter-stellar medium (ISM) were modest because of
effective matter mixing \citep[or if more massive CCSNe had greater iron
yields,][] {Wehmeyer2015}.

It should be noted that the aforementioned GCE models assume a single
halo system \citep[except for][]{Komiya2014, vandeVoort2015}, which is
incompatible with the currently dominant paradigm of the hierarchical
merging scenario for galaxy formation. \citet{Prantzos2006,
Prantzos2008} has shown that the overall shape of the Galactic halo's
metallicity distribution (MD) can be well reproduced by a merging
sub-halo model, with smaller sub-halos having suffered larger outflows
\citep[see also][]{Komiya2011}. More importantly for the issue of
$r$-process, \citet{Prantzos2006} showed that if the sub-halos evolved
at different rates, there would be no more a unique relation between
time and metallicity; in that case, he argued that the observed
``early'' (in terms of metallicity) appearance of $r$-elements and their
large dispersion can be explained, even if the main source of those
elements is NSMs.

In this Letter, we study the role of NSMs in the early Galaxy with a GCE
model based on the framework of such a hierarchical merging scenario and
on recent estimates of the progenitor binary lifetimes.

%%======================================================================
\section{Chemical Evolution of Sub-Halos}
%%======================================================================

\begin{deluxetable}{cccccc}
\tablecaption{Models of Sub-halos for Cases 1,2}
\tablewidth{0pt}
\tablehead{
\colhead{} \vspace{-1mm} & \colhead{} &  
\multicolumn{2}{c}{$k_{\rm OF}$ (Gyr$^{-1}$)} &
\multicolumn{2}{c}{$k_{\rm SF}$ (Gyr$^{-1}$)} \\ 
\colhead{$M_*$ ($M_\odot$)} \vspace{-1mm} &
\colhead{$\eta$} & 
\colhead{} & \colhead{} & \colhead{} & \colhead{} \\ 
\colhead{} & \colhead{} & 
\colhead{Case 1} & \colhead{Case 2} & 
\colhead{Case 1} & \colhead{Case 2}  
}
\startdata
 $10^4$         & 79  & 1.0 & 16   & 0.013 & 0.20 \\
 $10^5$         & 40  & 1.0 & 7.9  & 0.025 & 0.20 \\
 $10^6$         & 20  & 1.0 & 4.0  & 0.050 & 0.20 \\
 $10^7$         & 10  & 1.0 & 2.0  & 0.10  & 0.20 \\
 $10^8$         & 5.0 & 1.0 & 1.0  & 0.20  & 0.20 \\
 $2\times 10^8$ & 4.1 & 1.0 & 0.81 & 0.25  & 0.20   
\enddata
\end{deluxetable}

Each sub-halo is modeled as a one-zone, well-mixed single system that is
losing gas by outflow (because of stellar winds, SN explosions and tidal
interactions). The GCE model used in this work is the same as that of
the ISM evolution of the Galactic halo in \citet{Ishimaru1999,
Ishimaru2004, Wanajo2006} but it differs in what follows. Each sub-halo
is composed of stars with final total stellar mass $M_*$, evolving
homogeneously (i.e., with its ISM well mixed at every time).  We
consider sub-halos with $M_*/M_\odot = 10^4, 10^5, 10^6, 10^7, 10^8$,
and $2 \times 10^8$. The heaviest mass is set to be about half the
estimated stellar mass of the Galactic halo, $\sim 4 \times 10^8
M_\odot$ \citep{Bell2008}. The lowest mass corresponds to the one of
recently discovered ultrafaint dwarf galaxies \citep{Kirby2008}.

Although we have no direct observational clues for the property of each
sub-halo, recent observations of MDs of dwarf galaxies in the Local
Group can provide some hints \citep{Helmi2006}.  It is known that [Fe/H]
at the peak of a MD, [Fe/H]$_\mathrm{peak}$, corresponds to effective
yield $y_\mathrm{eff}$, i.e., the iron productivity of a galaxy system
which depends on the adopted stellar yields and IMF
\citep{Helmi2006,Prantzos2008}.
If a dwarf galaxy loses a significant amount of iron through outflow
with the rate (OFR) proportional to the star formation rate (SFR),
$y_\mathrm{eff}$ is approximately proportional to $\eta^{-1}$, where
${\rm OFR} \equiv \eta {\rm SFR}$ \citep{Prantzos2008}.
In practice, limited data of MDs are available and thus the observed
median metallicities ([Fe/H]$_0$) of dwarf galaxies are used instead of
[Fe/H]$_\mathrm{peak}$.  Observed mass-metallicity relation of dwarf
galaxies suggests that [Fe/H]$_0$ approximately scales as $M_*^{0.3}$
\citep{Kirby2013}.  We thus assume $y_\mathrm{eff}^{-1} \propto \eta =
5.0\, (M_*/10^8 M_\odot)^{-0.3}$ as in \citet{Prantzos2008}
\footnote{The values of the coefficient and the index are updated from
\citet{Prantzos2008} by adopting the latest mass-metallicity relation
\citep{Kirby2013}.}.
This can be naturally interpreted as a result of smaller SFR and/or
greater OFR because of the shallower gravitational well for a less
massive sub-halo (see also \S4).
SFR is assumed to be proportional to the mass of ISM ($M_{\rm ISM}$):
${\rm SFR} = k_{\rm SF} M_{\rm ISM}$, where the constant $k_{\rm SF}$
corresponds to the star formation efficiency, and ${\rm OFR} = k_{\rm
OF} M_{\rm ISM}$ as well, where $k_{\rm OF} = \eta k_{\rm SF}$.
However, we cannot constrain either of $k_{\rm SF}$ or $k_{\rm OF}$ from
observations.  Thus, we assume two extremes: Cases~1 and 2 for the fixed
ratios of $\eta$, respectively (Table~1).  The sub-halo of $M_* = 10^8
M_\odot$ is (arbitrary) set to have the same values of $k_{\rm SF}$ and
$k_{\rm OF}$ for both Cases~1 and 2.

The CCSN abundances of Mg and Fe (which we assume to be representative
of CCSN products and metallicity, respectively) are taken from
\citet{Nomoto2006}.  We ignore the Fe contribution from Type~Ia SNe,
because they appear to have played negligible role during the Milky Way
halo evolution (the observed $\alpha$/Fe ratio of halo stars is
$\sim$constant). The evolution of each sub-halo is computed up to 2~Gyr,
which is considered a reasonable estimate of the time-scale of the Milky
Way halo formation suggested from the observed ages of globular clusters
\citep{Roediger2014}.  The value of $k_{\rm SF}$ for the sub-halo with
$M_* = 10^8 M_\odot$ is set to account for the most metal rich stars in
the Galactic halo.

\begin{deluxetable}{ccccc}
\tablecaption{Parameters Adopted for NSMs}
\tablewidth{0pt}
\tablehead{
\colhead{Type} &
\colhead{$t_\mathrm{NSM}$ (Myr)} &
\colhead{$f_\mathrm{NSM}/f_\mathrm{CCSN}$} &
\colhead{$M_{\mathrm{Eu, NSM}}$ ($M_\odot$)}
}
\startdata
short-lived & 1.00 & $1.0\times 10^{-4}$ & $2\times 10^{-5}$  \\
long-lived  & 100  & $1.9\times 10^{-3}$ & $2\times 10^{-5}$
\enddata
\end{deluxetable}

Binary population synthesis models indicate an average binary lifetime
of $\langle t_\mathrm{NSM} \rangle \sim 1$~Gyr
\citep[e.g.,][]{Dominik2012}, being comparable with the lifetimes of
sub-halos. A significant fraction of NSMs are, however, expected to have
$t_\mathrm{NSM} \lesssim 100$~Myr \citep[$\sim
40\%$,][]{Belczynski2008}. In addition, a short-lived channel of
$t_\mathrm{NSM} \lesssim 1$~Myr is also predicted \citep{Belczynski2001,
Belczynski2002, DeDonder2004}. \citet{Dominik2012} estimate the fraction
of such a putative short-lived channel to be up to $\sim 7\%$ (for the
solar metallicity case). For illustrative purposes, we assume here a
bimodal distribution of $t_\mathrm{NSM} = 1$~Myr and 100~Myr and with
the corresponding fractions of 5\% and 95\%, respectively (Table~2).
For the total average frequency of NSMs ($f_\mathrm{NSM}$) relative to
that of CCSNe we assume $ \langle f_\mathrm{NSM} \rangle = 2\times
10^{-3} \langle f_\mathrm{CCSN} \rangle$ adopting the estimates of the
recent population synthesis calculations \citep{Dominik2012}. The mass
of the ejected Eu, representative of $r$-process elements in the
Galactic halo, is taken from the latest nucleosynthesis result of
\citet{Wanajo2014}, $M_{\mathrm{Eu, NSM}} = 2 \times 10^{-5} M_\odot$
\citep[see also][]{Goriely2011, Korobkin2012}. The mass of the ejected
Ba, which is also predominantly produced by the $r$-process in the early
Galaxy, is set to be $M_{\mathrm{Ba, NSM}} = 9\, M_{\mathrm{Eu, NSM}}$
according to the solar $r$-process ratio of these elements
\citep{Burris2000}.

%%======================================================================
\section{Enrichment of \lowercase{$r$}-Process Elements}
%%======================================================================

\begin{figure*}
%
%\epsscale{1.0}
\begin{center}
\includegraphics[width=0.42\textwidth, angle=-90]{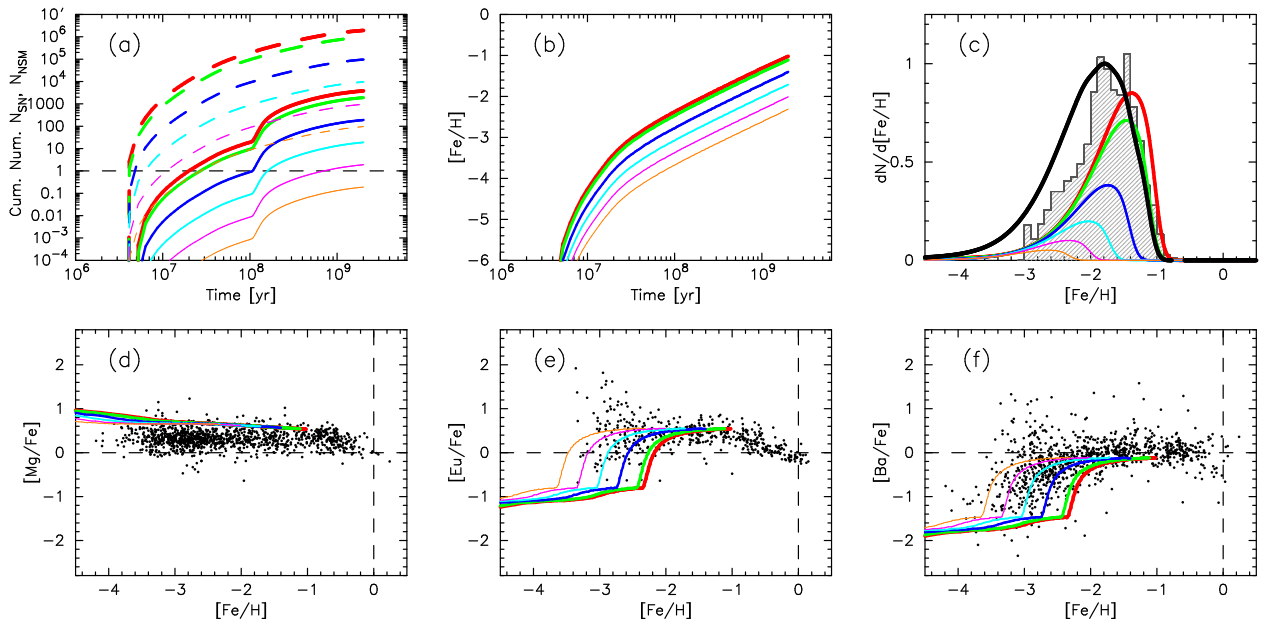}
\end{center}
\vspace*{5mm}
%\plotone{f1.eps}
%
\caption{Evolution of the sub-halos with $M_*/M_\odot = 10^4$, $10^5$,
$10^6$, $10^7$, $10^8$, and $2 \times 10^8$, respectively, indicated by
the thinnest to thickest curves for Case~1.  (a) Cumulative numbers of
NSMs (solid) and CCSNe (dashed) as functions of time. The horizontal
dashed line marks the number of unity (see text for implications). (b)
[Fe/H] temporal evolutions.  (c) MDs of sub-halos weighted with the
sub-halo mass function and their sum (thick-black). Observational data
of the Galactic halo (gray-hatched histogram) are taken from the
calibration catalog of \citet{An2013}.  (d)-(f) [Mg/Fe], [Eu/Fe], and
[Ba/Fe] as functions of [Fe/H], respectively.  The horizontal and
vertical lines indicate the solar values.  Observational stellar values
(dots) are taken from the SAGA database \citep{Suda2008}, excluding
carbon-enhanced stars that may have been affected by gas transfer in
binaries.  }  \label{fig:case1}
\end{figure*}

\begin{deluxetable}{ccccc}
\tablecaption{Results Related to MDs of Sub-halos}
\tablewidth{0pt}
\tablehead{
\colhead{} \vspace{-1mm} & 
\multicolumn{2}{c}{$M_D$ ($M_\odot$)} & 
\multicolumn{2}{c}{[Fe/H]$_{\rm peak}$} \\
\colhead{$M_*$ ($M_\odot$)} \vspace{-1mm} &
\colhead{} & \colhead{} & 
\colhead{} & \colhead{} \\ 
\colhead{} & 
\colhead{Case 1} & \colhead{Case 2} & 
\colhead{Case 1} & \colhead{Case 2} 
}
\startdata
 $10^4$         & $7.6 \times 10^6$ & $6.6 \times 10^6$    & $-2.63$ & $-2.56$\\
 $10^5$         & $3.8 \times 10^7$ & $3.3 \times 10^7$    & $-2.33$ & $-2.30$\\
 $10^6$         & $1.9 \times 10^8$ & $1.7 \times 10^8$    & $-2.03$ & $-2.02$\\
 $10^7$         & $1.0 \times 10^9$ & $9.0 \times 10^8$    & $-1.74$ & $-1.74$\\
 $10^8$         & $5.3 \times 10^9$ & $5.3 \times 10^9$    & $-1.46$ & $-1.46$\\
 $2\times 10^8$ & $8.8 \times 10^9$ & $9.3 \times 10^9$    & $-1.38$ & $-1.39$  
\enddata
\end{deluxetable}

Our results for Case~1 (with constant $k_{\rm OF}$) are shown in
Figure~\ref{fig:case1}. Those results for less (more) massive sub-halos
are indicated by the thinner (thicker) curves.  The cumulative number of
CCSNe in each sub-halo monotonically increases with time
(Figure~\ref{fig:case1}a). For NSMs, the cumulative number steeply rises
when the long-lived binaries start contributing at 100~Myr. [Fe/H]
monotonically increases with time for all the sub-halos; at a given time
[Fe/H] is greater for more massive sub-halos because of their higher
$k_{\rm SF}$.
MDs for all the sub-halos are also shown in Figure~\ref{fig:case1}c.
The metallicity at the peak of each MD, [Fe/H]$_{\rm peak}$ (Table~3),
is in agreement with the observed mass-metallicity relation
\citep{Kirby2013} scaled downward by $-0.4$ dex to exclude SNe Ia
contribution.
The dark halo mass of each sub-halo, $M_\mathrm{D}$ (Table~3), estimated
from the initial baryonic mass and the initial baryon to dark mass ratio
(which is assumed to be equal to the cosmic value $\Omega_\mathrm{B} /
\Omega_\mathrm{D} = 0.046/0.24 \sim 0.19$) implies $M_\mathrm{D} \propto
{M_*}^{0.7}$.  Since the reasonable mass function of $M_\mathrm{D}$ can
be regarded as $dN/dM_\mathrm{D} \propto M_\mathrm{D}^{-2}$
\citep[e.g.,][and the references therein]{Prantzos2008}, we obtain the
sub-halo mass function as $\propto M_*^{-1.7}$.  We find that the total
MD (thick-black curve) weighted with the sub-halo mass function is in
reasonable agreement with the observed one \citep[gray-hatched
region,][]{An2013}. We also find that the evolution of Mg
(Figure~\ref{fig:case1}d, representative of CCSN products) is in
reasonable agreement with the observed stellar abundances of Galactic
halo stars.
In contrast to $r$-process elements, the observed scatters in
[$\alpha$/Fe] is known to be as small as the measurement errors
\citep[e.g.,][]{Cayrel2004}.  This result is consistent with such small
scatters in [Mg/Fe], because each evolutionary trend is almost
independent of $M_*$.

The resulting evolution of Eu (representative of $r$-process elements)
is presented in Figure~\ref{fig:case1}e and compared to observed stellar
values. We notice a transition from slow to rapid evolution of [Eu/Fe]
for each sub-halo at $\sim$100 Myr, as a result of the contribution from
long-lived binaries starting. The corresponding [Fe/H] (see
Figure~\ref{fig:case1}b) differs from one sub-halo to another, being
lower than [Fe/H] $\sim -3$ for $M_* \le 10^6\, M_\odot$. This indicates
that the presence of stars at [Fe/H] $\sim -3$ with star-to-star scatter
in [Eu/Fe] ($\lesssim 0.5$) can be interpreted as a result of NSM
activity in sub-halos with various $k_{\rm SF}$.  Our model cannot,
however, explain the presence of $r$-process-enhanced stars with [Eu/Fe]
$> 1$. In addition our result predicts the presence of stars having
[Eu/Fe] $\sim -1$ for [Fe/H] $\lesssim -3$. This is due to the
contribution of the short-lived NSM at early times ($<
100$~Myr). Measurements of Eu in such stars would be challenging because
of the weak spectral lines. Such a signature has been seen in the Ba
abundances of EMP stars, [Ba/Fe] $\sim -2$ -- $-1$ at [Fe/H] $\lesssim
-3$ (Figure~\ref{fig:case1}f), which could also be explained as due to
the contribution of short-lived binaries. Note that the [Ba/Fe] values
for [Fe/H]$\gtrsim -2.5$ are underpredicted compared to the observed
ones; in this metallicity, contribution from the $s$-process becomes
important \citep{Burris2000}.

\begin{figure*}
%\epsscale{1.0}
\begin{center}
\includegraphics[width=0.42\textwidth, angle=-90]{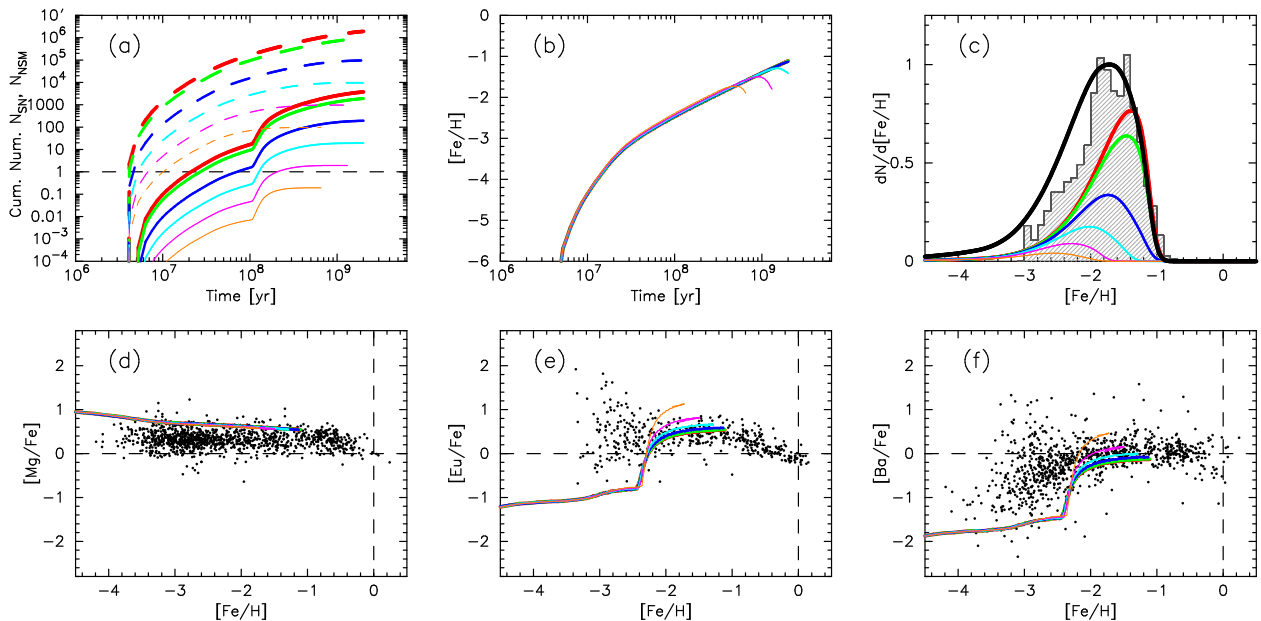}
\end{center}
\vspace*{5mm}
%\plotone{f2.eps}
\caption{Same as Figure~\ref{fig:case1}, but for Case~2.}
\label{fig:case2}
\end{figure*}

%\vspace*{5mm}

Figure~\ref{fig:case2} shows the results for Case~2, where a constant
$k_{\rm SF}$ is assumed. We find the cumulative numbers of NSMs and
CCSNe being similar to those in Case~1
(Figure~\ref{fig:case1}a). However, the evolutions stop earlier for less
massive sub-halos. This is due to the termination of star formation
because of gas removal by their significant outflows with greater values
of $k_{\rm OF}$.  In contrast to Case~1, the [Fe/H] evolution
(Figure~\ref{fig:case2}b) is identical among sub-halos because of the
same $k_{\rm SF}$, although the termination points differ depending on
their $k_{\rm OF}$. The resulting MDs (Figure~\ref{fig:case2}c) as well
as the evolutions of Mg (Figure~\ref{fig:case2}d) are similar between
Cases~1 and 2. In Figure~\ref{fig:case2}e, we find that the $r$-process
abundances increase to values greater than [Eu/Fe] $\sim 1$ for the
sub-halos with $M_* \le 10^5\, M_\odot$. This is a consequence of the
fact that the higher $k_{\rm OF}$ lead to smaller Fe amounts and thus
higher Eu/Fe ratios. It is interesting to note that, even without
invoking inhomogeneity of ISM, the enhancement of Eu can be in part
explained by our sub-halo models. However, our result here cannot
account for the $r$-process enrichment at [Fe/H] $ \sim -3$ because the
[Eu/Fe]'s start rising at [Fe/H] $\sim -2.4$ for all the sub-halos.

Our results imply that the reality may be between these two extremes,
Cases ~1 and 2; reasonable combinations of $k_{\rm SF}$ and $k_{\rm OF}$
might account for the presence of $r$-process-enhanced stars at [Fe/H]
$\sim -3$. It is also important to note that the cumulative numbers for
the least massive sub-halos ($M_* = 10^4\, M_\odot$) are $\sim 0.1$
around the end of their evolutions. This could be another source of
large enhancements of [Eu/Fe] ($\gtrsim 1$). The [Eu/Fe] values would be
substantially higher than the \textit{averaged} curves of our models,
provided that only a fraction of sub-halos experienced NSMs.

%%======================================================================
\section{Summary and Discussion}
%%======================================================================

We studied the role of NSMs for the chemical evolution of $r$-process
elements in the framework of the Galactic halo formation from merging
sub-halos. It is found that the appearance of Eu at [Fe/H] $\sim -3$
with star-to-star scatter in [Eu/Fe] ($\lesssim 0.5$) at [Fe/H] $\sim
-3$ can be interpreted as a result of lower star formation efficiency
$k_{\rm SF}$ for less massive sub-halos. On the other hand, the presence
of highly $r$-process-enhanced EMP stars ([Eu/Fe] $\gtrsim 0.5$) can be
explained if values of $k_{\rm OF}$
(the multiplicative factor for the outflow rate)
are higher for less massive sub-halos.
These may be reasonable assumptions because less massive sub-halo
systems have weaker gravitational potential (as indicated by
$M_\mathrm{D}$ in Table~3) and thus are expected to form stars less
efficiently and/or expel the ISM more easily.
The ratio of OFR to SFR, $\eta$, is assumed to be proportional to
${M_*}^{-0.3}$. 
Recent observations of relatively massive galaxies ($10^7 - 10^{11}
M_\odot$) also suggest similar anti-correlation between $M_*$ and $\eta$
\citep{Chisholm2014}.
Under this assumption, the metallicity at the peak of each sub-halo's MD
(Table~3) appears to be consistent with the observed mass-metallicity
relation.  In addition, the total MD weighted with the sub-halo mass
function shows reasonable agreement with the observed one of the
Galactic halo.

The low-level Ba abundances ([Ba/Fe] $\sim -1.5$) observed in the EMP
stars of [Fe/H] $\lesssim -3$ may be due to the contribution from the
short-lived binaries (with $t_\mathrm{NSM} = 1$~Myr in this
study). Thus, the main features of $r$-process abundances observed in
EMP stars, their appearance at [Fe/H] $\sim -3$ with large star-to-star
scatter and their sub-solar amounts for [Fe/H] $\lesssim -3$, can be
potentially explained by the contribution from solely NSMs.

The highly Eu enhanced stars at [Fe/H]$\sim -3$ may be accounted by
reasonable combinations of $k_{\rm SF}$ and $k_{\rm OF}$.  The small
cumulative numbers of NSMs ($<1$) for least-massive sub-halos could be
additional reason for the observed large scatter in [Eu/Fe], because
NSMs may occur only in some of them, while no NSMs occur in others.
Once NSMs occur in such less massive systems, their ISM must be highly
enriched by $r$-process elements.

In conclusion, our results imply that the current observational
evidences of $r$-process signatures in EMP stars can be interpreted as a
consequence of the Galactic halo formation from merging sub-halos. If
the model is a reasonable simplification for such chemo-dynamical
evolutions of sub-halos, NSMs can be the main $r$-process contributors
throughout the Galactic history. Contributions from CCSNe are not
necessarily needed as invoked in previous GCE studies.
Note that our assumption of well-mixed ISM might not necessarily be an
oversimplification, at least for less-massive sub-halos, in which a
small number of explosive events would easily homogenize their smaller
amount of ISM before the next episode of star formation.
The observed small scatters of abundance ratios of other elements such
as iron-peak elements or $\alpha$-elements in EMP stars should be
examined using a consistent assumption.
Possible effects of ISM inhomogeneity as well as of stochastic nature of
NSM events (i.e., $N_{\rm NSM}<1$) will be discussed in a forthcoming
paper.

%%======================================================================
\acknowledgements
%%======================================================================

This work was supported by the RIKEN iTHES Project and the JSPS
Grants-in-Aid for Scientific Research (26400232, 26400237).

%%======================================================================
% References
%%======================================================================

\clearpage

%%======================================================================
\end{document}